\documentstyle[12pt,epsfig]{article}

\newcommand{\be}{\begin{eqnarray}}
\newcommand{\ee}{\end{eqnarray}}
\newcommand{\etal}{{\it et al.}}
\def\nue{{\nu_e}}

\newcommand{\dm}{\mbox{$\Delta m_{21}^2$~}}

\newcommand{\kl}{\mbox{KamLAND~}}

\newcommand{\thsol}{\mbox{$\theta_{12}$~}}

\def\ltap{\ \raisebox{-.4ex}{\rlap{$\sim$}} \raisebox{.4ex}{$<$}\ }

\newcommand{\sss}{\sin^2 \theta_{12}}
\begin{document}

\begin{flushright}
SISSA 78/2003/EP     \\
SINP/TNP/03-32\\
hep-ph/0309174
\end{flushright}

\begin{center}
{\Large \bf Constraints on neutrino oscillation parameters   
from the SNO salt phase data}
\vspace{.5in}

{ Abhijit Bandyopadhyay$^1$,
Sandhya Choubey$^{2,3}$,
Srubabati Goswami$^{4,5}$,
S.T. Petcov$^{3,2,6}$,
D.P. Roy$^7$}
\vskip .5cm

$^1${\small {\it Saha Institute of Nuclear Physics}},
{\small {\it 1/AF, Bidhannagar,
Calcutta 700 064, India}},\\

$^2${\small {\it INFN, Sezione di Trieste, Trieste, Italy}},\\
$^3${\small {\it Scuola Internazionale Superiore di Studi Avanzati,
I-34014,
Trieste, Italy}},\\
$^4${\small {\it Harish-Chandra Research Institute, Chhatnag Road, Jhusi,
Allahabad  211 019, India}},\\
$^5${\small{\it The Abdus Salam International Centre for Theoretical Physics,
I-34100, Trieste, Italy}},\\
$^6${\small {\it Institute of Nuclear Research and Nuclear Energy, 
Bulgarian Academi of Science, Sofia, Bulgaria}},\\
$^7${\small {\it Tata Institute of Fundamental Research, Homi Bhabha Road, 
Mumbai 400 005, India}}

\vskip 1in

\end{center}
\begin{abstract}
The physics implications of
the just published salt phase data from 
the SNO experiment are examined.
The effect of these data on the allowed ranges of the
solar neutrino oscillation parameters, \dm and $\sss$,
are studied in the cases of two- and three- neutrino
mixing. In the latter case we derive an upper limit
on the angle $\theta_{13}$.
Constraints on the solar $\nu_e$ 
transitions into a mixture of active and sterile neutrinos
are also presented. Finally, we give
predictions for the day-night asymmetry 
in the SNO experiment, 
for the event rate in the BOREXINO and LowNu experiments,
and discuss briefly the constraints on 
the solar neutrino oscillation parameters
which can be obtained with prospective KamLAND data.
\end{abstract}

\newpage

\section{Introduction}
\vspace{-0.3cm}

The past two years  have witnessed 
remarkable experimental progress 
in the studies of neutrino mixing and oscillations.
The latest addition to this magnificent effort is the 
salt phase data from the SNO experiment \cite{Ahmed:2003kj}.
 
In 2001, the evidences for solar neutrino oscillations
\cite{Pont67} obtained 
in the pioneering experiment of Davis et al. (Homestake) 
\cite{Pont46,Davis68},
and in the Kamiokande, SAGE, GALLEX/GNO \cite{sol} and Super-Kamiokande
\cite{SKsolar}
experiments, were reinforced by the first results 
of the SNO experiment on the charged current (CC)
reaction on deuterium induced by solar neutrinos \cite{Ahmad:2001an}. 
In conjunction with the Super-Kamiokande (SK)
$\nu - e^-$ scattering data, the SNO CC data
established the existence of solar $\nu_e$ flavour 
conversion with a statistical significance of 3.3$\sigma$.
This conclusion was further corroborated by
the 2002 SNO data on the neutral current (NC) 
reaction on deuterium, caused by solar neutrinos \cite{Ahmad:2002jz}. 
The combined CC and NC SNO data 
showed at 5.3$\sigma$ the presence
of a nonzero 
$\nu_{\mu,\tau}$ and/or $\bar{\nu}_{\mu,\tau}$ 
component in the flux of the solar $^8B$  
neutrinos reaching the Earth.
At each stage, the SNO data enabled one to determine 
with improved precision
the average solar $\nu_e$ survival probability 
$P_{ee}$ from the CC reaction data, 
and the $^{8}{B}$ flux normalisation $f_B$ from the 
data on the NC reaction.
This in turn led to a diminishing of 
the allowed regions of values of the two
parameters - the neutrino mass squared difference 
$\dm$ ($\equiv \Delta m_\odot^2$) and the mixing angle 
$\thsol$ ($\equiv \theta_{\odot}$),
characterizing the solar neutrino oscillations. 
The SNO CC data had the main impact of ruling out the 
Small Mixing Angle (SMA) MSW \cite{msw} 
solution in conjunction with the SK data
\cite{Ahmad:2001an,snocc,snoccothers}.
The SNO data from the $D_2O$ phase\footnote{Here, and in the rest of 
the paper, by $D_2O$ phase we mean the NC reactions due to 
the final state neutron capture on deuterium.}
showed a clear preference for the 
Large Mixing Angle (LMA) MSW solution of the solar neutrino
problem, disfavoring 
the relatively low $\dm$ 
(LOW, QVO) solutions \cite{Ahmad:2002jz,snonc}.
The first results of the KamLAND experiment \cite{KamLAND},
under the plausible assumption of CPT-invariance
in the lepton sector, established the
LMA solution as unique solution of the 
solar neutrino problem. 

The combined two-neutrino oscillation 
analyses of the 
solar neutrino and KamLAND  data, available
before the publication of the SNO salt phase results,
identified two distinct solution sub-regions within 
the LMA solution region at 99\% C.L.~ 
(see, e.g., \cite{solfit1,solfit2}).
The best fit values of 
$\Delta m_\odot^2 $ and $\theta_\odot$
in the two sub-regions - 
low-LMA and 
high-LMA,
were found to be 
$\Delta m_\odot^2 = 7.2 \times 10^{-5}~{\rm eV^2}$, 
$\sin^2 \theta_\odot = 0.3$, 
and 
$\Delta m_\odot^2 = 1.5 \times 10^{-4}~{\rm eV^2}$, 
$\sin^2 \theta_\odot = 0.3$, respectively \cite{solfit1}.
The low-LMA solution was preferred 
statistically by the data. 
At 99.73\% C.L. 
(3$\sigma$) the two regions merged and
one obtained: 
\begin{equation}
\Delta m_\odot^2 \cong 
(5.0 - 20.0) \times 10^{-5}~{\rm eV^2},~~~ 
\sin^2 \theta_\odot \cong (0.21 - 0.47)~.
\label{sol90}
\end{equation}
%

 In the case of 3-neutrino mixing, the
analysis of the solar neutrino and KamLAND data
involves an additional parameter $\theta_{13}$ - the
neutrino mixing angle limited by the CHOOZ and Palo
Verde experiments \cite{CHOOZPV}. 
The precise upper limit on $\sin^2\theta_{13}$
depends on the value of $\Delta m^2_{31}$ - the neutrino 
mass squared difference responsible for the atmospheric
$\nu_{\mu}$ and $\bar{\nu}_{\mu}$ oscillations. 
The preliminary results of an improved analysis of the
SK atmospheric neutrino data, performed recently by the
SK collaboration, gave \cite{SKatmo03}
\begin{equation} 
1.3 \,\times\, 10^{-3}\,\mbox{eV}^2\,\ltap\,
\Delta m^2_{31}\, \ltap\,3.1\, \times\,10^{-3}\,\mbox{eV}^2~,
 ~~~~~90\%~{\rm C.L.},
\label{atmo03}
\end{equation}
%
\noindent with a best fit value $\Delta m^2_{31} = 2.0\times 10^{-3}$ eV$^2$.

   After the \kl results 
the two major issues to be settled with the future
solar neutrino and KamLAND data are:

\begin{itemize}
\item Resolving the ambiguity between the low-LMA and the high-LMA 
solutions and thereby obtaining tighter constraints on \dm. 

\item Constraining further the allowed range of the solar neutrino 
mixing angle, $\theta_{12}$.
\end{itemize}

  In the present article we examine 
some of the physics implications of
the latest salt phase data from SNO 
\cite{Ahmed:2003kj}. We study, in particular,
the effect of these data on the allowed ranges of the
solar neutrino oscillation parameters, \dm and $\sss$.
This is done in the cases of two- and three- neutrino
mixing. In the latter case we obtain an upper limit
on the angle $\theta_{13}$.
Constraints on the solar $\nu_e$ 
transitions into a mixture of active and sterile neutrinos,
i.e., on the allowed sterile fraction, 
are also presented. Finally, we give
predictions for the day-night asymmetry 
in the SNO experiment, 
for the event rate in the BOREXINO and LowNu experiments,
and discuss briefly the constraints on 
the solar neutrino oscillation parameters
which can be obtained with future KamLAND data.

\vspace{-0.4cm}
\section{Two-neutrino oscillation analysis}
\vspace{-0.3cm}
\subsection{Analysis with global solar neutrino data}

\begin{figure}[t]
\centerline{\psfig{figure=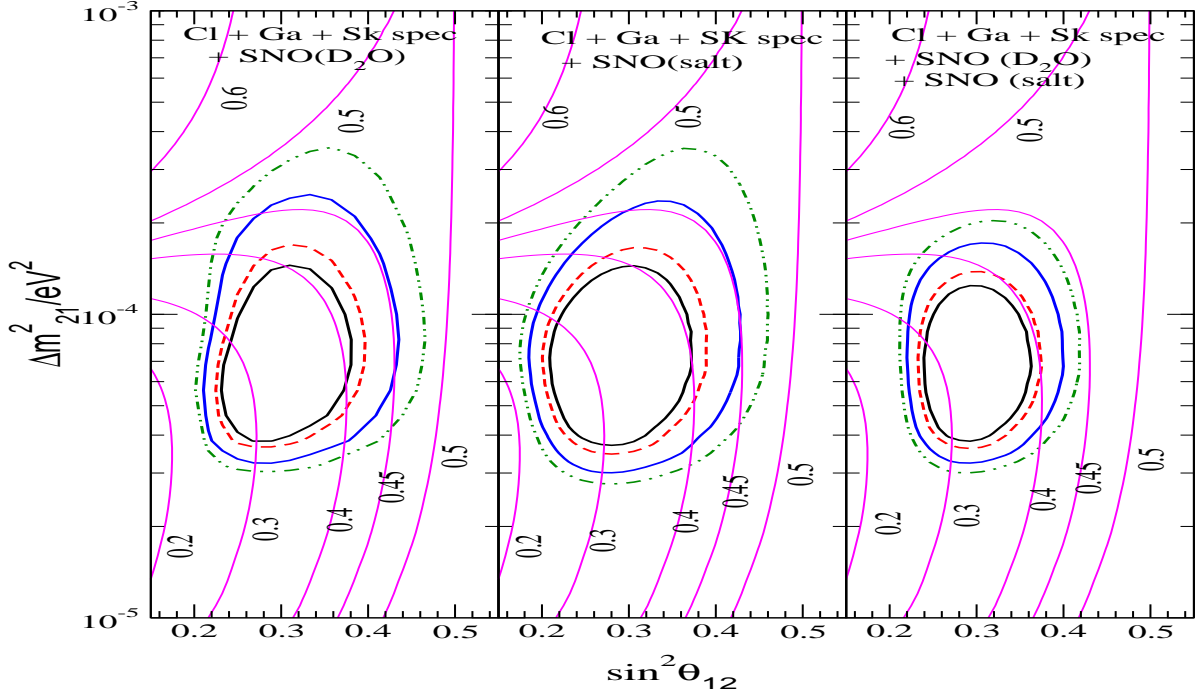,height=5.in,width=7.0in}}
\caption{The 90\%, 95\%, 99\% and 99.73\% C.L. 
allowed regions in the $\dm-\sss$ plane from global 
$\chi^2$-analysis of the data from solar neutrino experiments.
We use the $\Delta \chi^2$ values 
corresponding to a two parameter fit
to plot the C.L. contours.
Also shown are the lines of constant CC/NC event rate 
ratio $R_{CC/NC}$.}
\label{sol2osc}
\end{figure}

\begin{table}
\begin{center}
\begin{tabular}{c|cc|cc}
\hline\hline
Data & \multicolumn{2}{c|} {best-fit parameters}  & \multicolumn{2} {c}
{99\% C.L. allowed range}\\
\cline{2-5}
set used & \dm/($10^{-5}$eV$^2$) & $\sss$ & \dm/($10^{-5}$eV$^2$) & $\sss$\\
\hline \hline
Cl+Ga+SK+$D_2O$& $6.06$ & 0.29 & $3.2-24.5$ &  $0.21 - 0.44$ \\
Cl+Ga+SK+salt& $6.08$ & 0.28 & $3.0-23.7$ &  $0.19 - 0.43$ \\
Cl+Ga+SK+$D_2O$+salt& $6.06$ & 0.29 & $3.2-17.2$ &  $0.22 - 0.40$ \\
Cl+Ga+SK+$D_2O$+KL& $7.17$ & 0.3 & $5.3-9.9$ &  $0.22 - 0.44$ \\
Cl+Ga+SK+$D_2O$+salt+KL& $7.17$ & 0.3 & $5.3-9.8$ &  $0.22 - 0.40$ \\
\hline \hline
\end{tabular}
\label{tab1}
\caption{
The best-fit values of the solar neutrino oscillation 
parameters, obtained using different combinations 
of data sets. Shown also are the 99\% C.L. (corresponding to $\Delta \chi^2$ 
for a 2 parameter fit) allowed ranges of the parameters from 
the different analyses.
}
\end{center}
\end{table}

In this Section we first perform a two-neutrino oscillation analysis of 
the global solar data, incorporating the new SNO results. 
We include the total rates from the radiochemical experiments 
Cl and Ga (Gallex, SAGE and GNO combined) \cite{sol} and 
the 1496 day 44 bin SK Zenith 
angle spectrum data \cite{SKsolar}
~\footnote{SK has recently  
reanalyzed their day/night 
data with improved precision \cite{SKDN}. 
However, the information content in 
\cite{SKDN} is not enough for including it in our analysis.}.
For SNO we take the combined CC, NC and Electron Scattering (ES)
34 bin energy spectrum data from the
$D_2O$ phase \cite{Ahmad:2002jz}, and the 
recently reported CC, NC and ES rates
from the latest salt phase of the experiment
\cite{Ahmed:2003kj} 
~\footnote{Note that the 
salt enriched data from SNO gives the 
CC, ES and NC total rates without any assumption on the 
$^8B$ spectrum shape.}.  
To ascertain the impact of 
the salt phase data and facilitate comparison
between the impact of the data from the different phases,
the SNO data is included in the following three ways in our
analysis:  
\begin{enumerate} 
\item Use only the CC+ES+NC day/night spectra data from the $D_2O$ phase, 
\item Use only the CC, ES and NC event rate data from the salt phase,
\item Use data from both phases together.
\end{enumerate} 
We follow the instructions 
given in \cite{howto} by the SNO Collaboration
in treating  the SNO data. 

For our statistical analysis of the global solar neutrino data 
we follow a covariant approach and 
minimise the $\chi^2$ defined as 
\be
\chi^2_{\odot} = \sum_{i,j=1}^N (R_i^{\rm expt}-R_i^{\rm theory})
(\sigma_{ij}^2)^{-1}(R_j^{\rm expt}-R_j^{\rm theory})
\label{chi2}
\ee
where $R_{i}$ are the solar data points, $N$ is
the number of data points and
$(\sigma_{ij}^2)^{-1}$ is the inverse of the covariance matrix,
containing the squares of the correlated and uncorrelated experimental
and theoretical errors. 
The $^8B$ flux normalisation factor 
$f_B$ is left to vary freely in the analysis. 
For further details of our solar neutrino data 
analysis we refer the reader
to our earlier papers \cite{snocc,snonc}.

The results of the analysis of the global solar neutrino data are
presented in Table 1 and Figure \ref{sol2osc}. Table 1 gives the best-fit 
points and the allowed range of parameter values.
The best-fit for the global analysis, including the complete 
SNO data from both phases,
is obtained at $\dm = 6.06 \times 10^{-5}$ eV$^2$,
$\sss = 0.29$ and $f_B=1.04$. 
Note that if we consider 
only the salt phase data from SNO, 
the best-fit value of $\sss$ is marginally lower.

In the left panel of Figure \ref{sol2osc}
we show the allowed region in the parameter space 
when only the spectrum data 
from the $D_2O$ phase are included. 
In the middle panel the allowed region,
obtained by including the  
SNO salt phase data but
excluding the SNO spectrum data from the $D_2O$ phase, is 
presented. 
 A comparison of the two panels 
shows that with the exclusion of the 
$D_2O$ phase spectrum data, the allowed region enlarges in size. 
Even though the SNO data from the  $D_2O$ phase agrees remarkably well 
with the salt phase data, the ratio of CC and NC event
rates, $R_{CC/NC}$, is slightly different 
for the two phases. In particular, for the $D_2O$ phase, if one uses the 
data given by SNO for the null hypothesis, one gets for 
the ratio 
$R_{CC/NC} = 0.346$.
For the salt phase, the ratio is $R_{CC/NC} = 0.306$.
Thus, the CC to NC event rate
ratio has decreased, which has very important
implications in constraining $\sin^2\theta$ and $\Delta m^2$ 
\cite{Gonzalez-Garcia:2000ve,maris} (see also \cite{ccnc}). 
In Figure \ref{sol2osc} 
the iso-$R_{CC/NC}$
contour lines 
are superimposed 
on the allowed regions
in the $\dm -\sin^2\theta_{12}$ plane. 
The figure clearly shows 
that since the $R_{CC/NC}$ for the 
salt phase data is lower, 
the allowed regions shift left, 
following the iso-$R_{CC/NC}$ contours.
This results in the shift of the allowed range
of $\sss$ to  smaller values (see Table 1).
The allowed \dm shift to 
lower values as well for the same reason. 

The third panel shows the allowed regions,
obtained by including the SNO data 
from both the $D_2O$ phase and the salt phase. 
Following the SNO collaboration \cite{howto}, we treat these two phases 
as separate experiments with no correlation between them. 
Combination of the salt phase and the $D_2O$ phase data produce a more
restrictive upper bound on \dm $\leq 1.7\times 10^{-4}$ eV$^2$ 
(99.73\% C.L.). The 
upper limit on $\sss$ also improves compared to what we have before the 
salt phase data, as can be seen by comparing the first panel with the 
last one in Figure \ref{sol2osc}. 
The exact intervals in which the solar 
neutrino oscillations parameters
are constrained to lie are given in Table 1.

\subsection{Constraints from combined solar and \kl data}
 
\begin{figure}[t]
\centerline{\psfig{figure=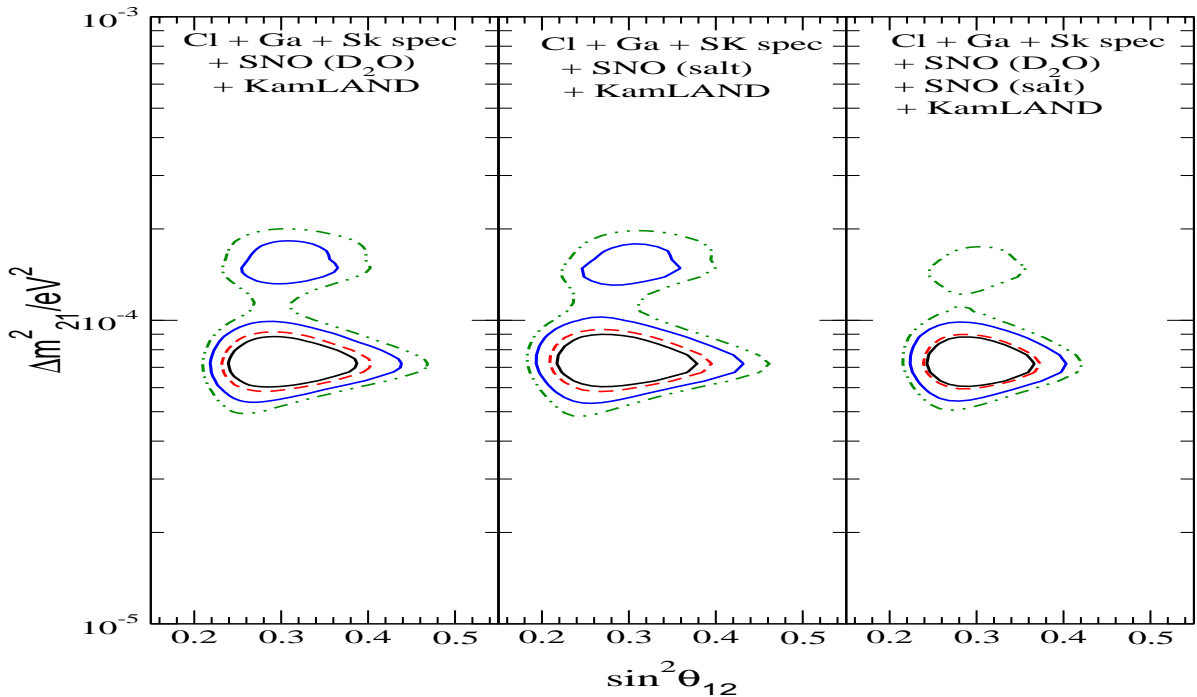,height=5.in,width=7.in}}
\caption{The 90\%, 95\%, 99\% and 99.73\% C.L.
allowed regions in the $\dm-\sss$ plane from global 
$\chi^2$-analysis of solar and \kl data.
We use the $\Delta \chi^2$ values corresponding to a
2 parameter fit 
to plot the C.L. contours.
}
\label{solkl2osc}
\end{figure}

\begin{figure}[t]
\centerline{\psfig{figure=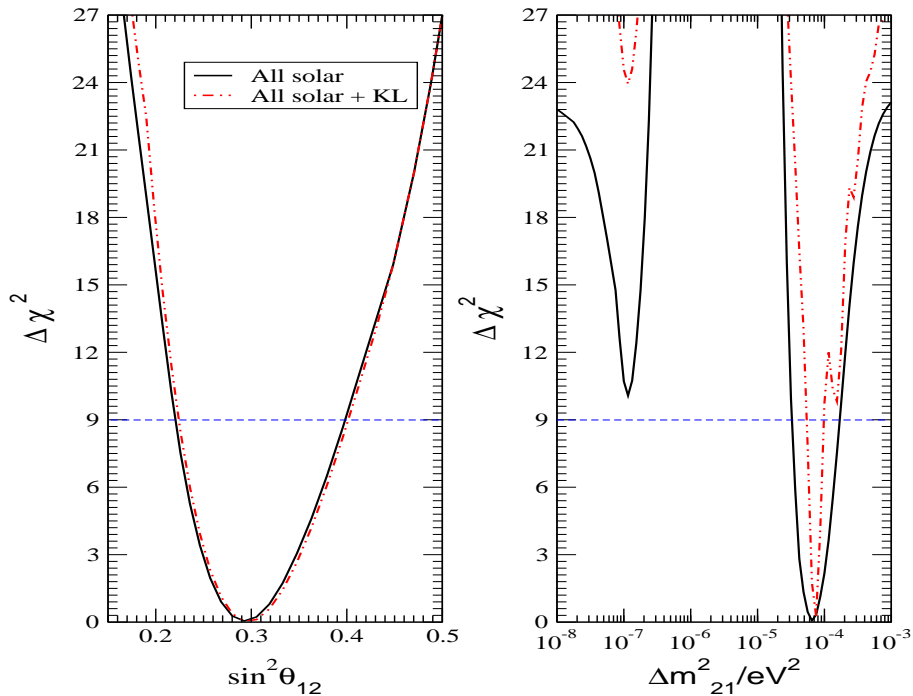,height=5.in,width=5.in}}
\caption{Bounds on \dm and $\sss$ from
the $\Delta \chi^2 $ as a function of \dm and $\sss$, respectively.
The results shown in both panels are obtained by allowing
all the other
parameters to vary freely.
The dashed line shows the $3\sigma$ limit corresponding
to 1 parameter fit.
}
\label{delvs12}
\end{figure}

We next include the 162 ton-year results from the \kl 
experiment in the analysis. We use the 13 bin
\kl spectrum data and defined a $\chi^2$ 
assuming a Poissonian distribution as
\be
\chi^2_{klspec}=
\sum_{i}\left[2(X_n S_{KL,i}^{theory} - S_{KL,i}^{expt}) 
+ 2 S_{KL,i}^{expt} \ln(\frac{S_{KL,i}^{expt}}
{X_n S_{KL,i}^{theory}})\right] + \frac{(X_n -1)^2}{\sigma^2_{sys}}
\label{chip}
\ee
where $\sigma_{sys}$ is taken to be 6.42\% and $X_n$ allowed to vary 
freely (see \cite{solfit1} for the details of the analysis).
In the last 2 rows of 
Table 1 we give the best-fit data points and the allowed ranges of 
the parameters obtained before and after including the latest salt 
phase SNO data in the global analysis. 
The best-fit point for the combined global analysis 
is in the low-LMA region at 
$\dm=7.17 \times 10^{-5}$ eV$^2$ and $\sss=0.30$. 
Thus, the best-fit values of the parameters 
for the combined solar+KamLAND data analysis 
do not change after inclusion of the latest SNO results.
The inclusion of the salt data results in an improvement in the 
precision of the $^{8}{B}$ flux normalisation from
$f_B = 1.01^{+0.19}_{-0.19}$, obtained 
without the salt phase data, 
to $f_B = 1.02^{+0.14}_{-0.15}$ 
(3$\sigma$ results). 
The agreement with the Standard Solar Model prediction \cite{bbp2000}
is excellent. 

In Figure \ref{solkl2osc} we present the 
corresponding allowed regions.
Again we show results separately for only the 
$D_2O$ phase (left-hand panel), only the salt phase (middle panel) 
and the global data, with the two phases combined (right-hand panel).
Note the shift of the allowed regions in the middle panel which 
includes only the salt phase data, to smaller values of $\sss$. 
We find that 
the separate inclusion of the data from each phase of SNO, in the 
combined analysis  with the \kl data,
allows the high-LMA solution at 99\% C.L..  
However when the global solar neutrino data -- with data 
from both phases 
of SNO combined -- is included with the \kl data, 
the high-LMA region is allowed
only at 99.13\% C.L. (2.63$\sigma$) 
corresponding to a difference of $\chi^2$ of 9.5, 
with respect to the 
global $\chi^2_{min}$ obtained in the low-LMA region.

\begin{figure}[h]
\centerline{\psfig{figure=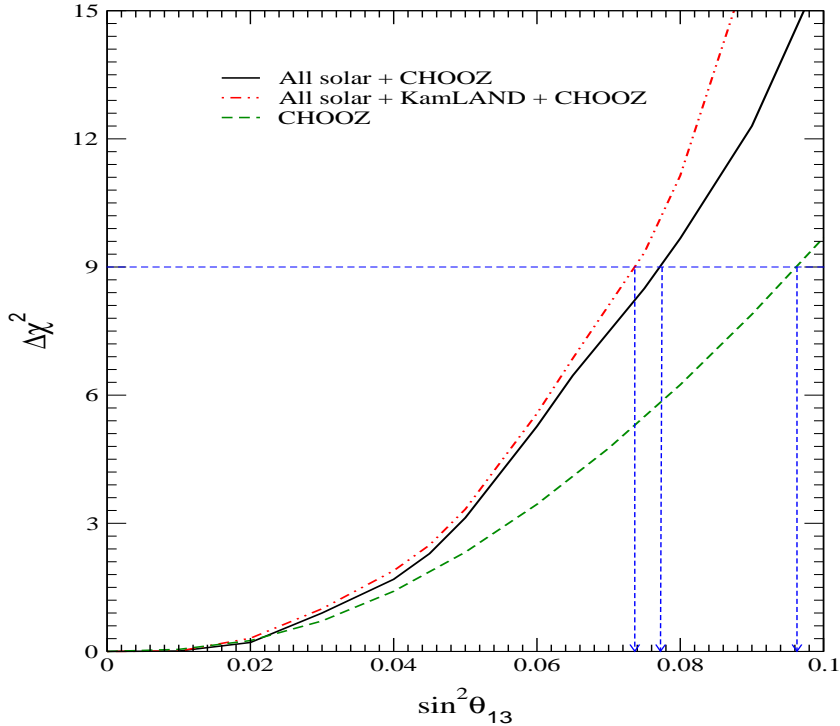,height=5.in,width=5.in}}
\caption{
Bounds on the mixing angle $\theta_{13}$ from
the CHOOZ data only (dashed line), the global solar neutrino 
and CHOOZ data (solid line), and the combined solar, CHOOZ and \kl 
data (dot-dashed line).
The $\Delta m^2_{31}$ is allowed to vary freely within the 3$\sigma$ range 
($2.0^{+1.2}_{-0.9} \times 10^{-3}$ eV$^2$),
which is obtained using the  updated 
SK and K2K results \cite{lisiatmupdate}.   
All the other parameters are allowed to vary freely.
The short-dashed line shows the $3\sigma$ limit corresponding to the case of
1 parameter fit.
}
\label{delchith13}
\end{figure}

In Figure \ref{delvs12} we show 
the dependence of $\Delta \chi^2 = \chi^2 - \chi^2_{min}$ on 
\dm and $\sss$ 
respectively, after marginalising over the remaining free parameters. 
The solid lines represent the $\Delta \chi^2$ 
for the global solar neutrino data, while the dot-dashed curves 
correspond to the combined solar and \kl data.
We note that the \dm corresponding to the 
LOW solution is ruled out at slightly  
more than $3\sigma$ by the solar neutrino 
data alone, and at 
nearly $5\sigma$ from 
the combined solar and \kl data. \kl results are seen to produce a 
remarkable constraint on $\dm$. 
Maximal mixing is now ruled out at more 
than $5\sigma$ level by the solar data alone. 
The \kl data does not 
put a strong constraint on $\sss$.

\vspace{-0.4cm}
\section{Bounds from three-neutrino oscillation analysis}
\vspace{-0.3cm}

\begin{figure}[p]
\centerline{\psfig{figure=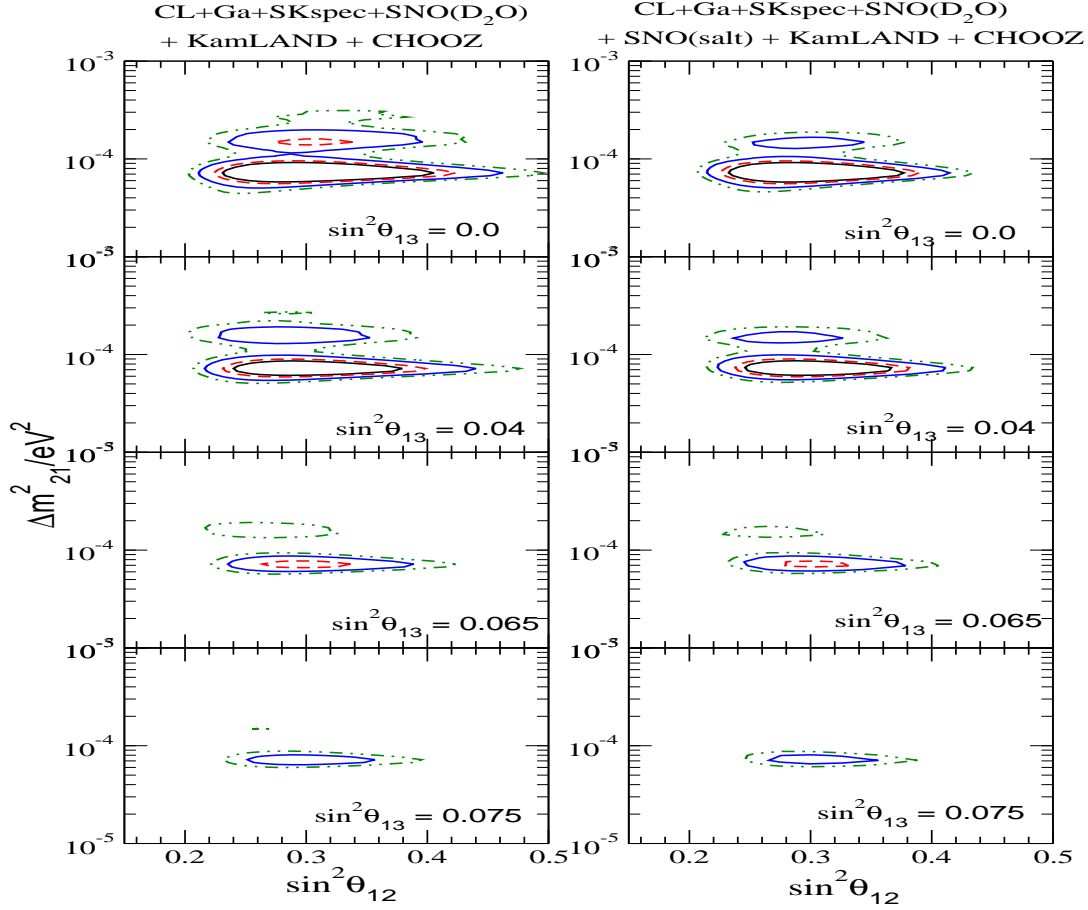,height=5.in,width=6.in}}
\caption{
The 90\%, 95\%, 99\% and 99.73\% C.L. allowed contours in the 
$\dm-\sss$ plane, from the three-neutrino oscillation analysis of the 
global solar and reactor data, including the data from the 
\kl and CHOOZ experiments. 
The 
different panels are drawn at 
different fixed values of $\sin^2\theta_{13}$. 
$\Delta m^2_{31}$ is allowed to vary freely within  
appropriate  
range taken from \cite{lisiatmupdate}.
The left hand panels are before the SNO salt data while the right hand panels
are including the SNO salt data.
Here we use three parameter fit 
$\Delta \chi^2$ values to plot the C.L. contours.
}
\label{solkl3osc}
\end{figure}

So far we have presented results obtained in the framework of 
two-neutrino oscillations,
where the solar $\nu_e$ 
oscillates into another active neutrino 
with a different flavor. However, 
if the mixing angle $\theta_{13}$
which is restricted by the CHOOZ and Palo Verde
data is not zero, 
the solar $\nu_e$ oscillations will
involve also the third heaviest neutrino mass eigenstate,
with the associated neutrino mass squared
difference given by $\Delta m^2_{31}$.
The relevant electron neutrino/antineutrino survival 
probability in the three-neutrino mixing
case is given by the following expression:
\be
P_{ee}^{3gen} \cong \cos^4 \theta_{13} P_{ee}^{2gen} + \sin^4\theta_{13}
\label{3genpee}
\ee
where $ P_{ee}^{2gen}$ is the $\nu_e$
survival probability for
two-neutrino mixing (see, e.g., \cite{SP3nuosc88}). 
In the case of solar neutrinos,
$ P_{ee}^{2gen} \equiv P_{ee\odot}^{2gen}$ 
is the two-neutrino oscillation
$\nu_e$ survival probability  \cite{SP88} with
the solar electron number density $N_e$
replaced by $N_e \cos^2\theta_{13}$.
The term $\sin^4\theta_{13}$ can 
be neglected 
to a good approximation 
in eq. (\ref{3genpee}).
Since for the solar neutrinos one has
$P_{ee\odot}^{2gen} \approx \sin^2\theta_{12}$ in the low-LMA 
region, we have $P_{ee\odot}^{3gen} 
\approx\cos^4 \theta_{13} \sin^2\theta_{12}$.
For the reactor $\bar{\nu}_e$ detected in the
\kl experiment, the 
matter effects are negligible and one gets
$P_{eeKL}^{3gen} = \cos^4 \theta_{13} \{1 - \sin^22\theta_{12}
\sin^2(\dm L/4E)\}$, where $L$ is the 
source-detector distance and $E$ the antineutrino energy.
As  Eq. (\ref{3genpee}) indicates,
the presence of a non-zero $\theta_{13}$ shifts
$\theta_{12}$ obtained from the two-neutrino oscillation
solar neutrino data analysis
to larger values. In contrast, 
for  the reactor antineutrinos, 
one can expect 
the allowed range of $\theta_{12}$, 
determined from a two-neutrino mixing analysis
of the reactor antineutrino data, 
to shift to smaller 
values, if $\theta_{13}$ is non-zero and is sufficiently large. 
The survival probability for the short-baseline CHOOZ experiment is 
approximately given by $P_{eeCHOOZ}^{3gen} \approx 1 - \sin^22\theta_{13}
\sin^2(\Delta m_{31}^2 L/4E)$ 
and is hence very sensitive to the 
range of allowed value of atmospheric neutrino mass squared difference, 
$\Delta m_{31}^2$ 
\footnote{In our numerical calculation we use the  probability keeping the 
$\Delta m^2_{21}$ terms \cite{BNPChooz}.}. 
Therefore with the shift of 
$\Delta m^2_{31}$  
to lower values in the new SK analysis,
the constraint on $\theta_{13}$ from CHOOZ 
is expected to get affected.
The information in \cite{SKatmo03} is not sufficient for a 
full global analysis using the SK atmospheric data. 
We adopt the following procedure. 
We allow $\Delta m^2_{31}$ to vary freely within  the 
range given in \cite{lisiatmupdate},   
obtained using 
the updated SK \cite{SKatmo03} and K2K \cite{k2k} results  
and 
perform a combined three-neutrino oscillation 
analysis of the global solar 
neutrino and  reactor
data, including both the \kl and CHOOZ results  
(for earlier global three generation analyses see, e.g. 
\cite{3gen1}).
 
In Figure \ref{delchith13} we present the $\Delta \chi^2$ obtained 
for various fixed values of $\sin^2\theta_{13}$, when all other 
parameters are allowed to vary freely. 
The $3\sigma$ bounds on 
$\sin^2\theta_{13}$, obtained from the global solar 
neutrino and CHOOZ data analysis,
can be directly read from the figure as 
$\sin^2\theta_{13} < 0.077$. 
The bound 
derived from the 
combined analysis of the solar neutrino, 
CHOOZ and KamLAND data
tightens somewhat to $\sin^2\theta_{13} < 0.074$.
\footnote{If we let $\Delta m^2_{31}$ 
to vary within the range allowed by the earlier SK analysis we get 
$\sin^2\theta_{13} < 0.057$. Thus the ratio 
of the old and the new bound on
$\sin^2\theta_{13}$ is 1.3 and this
matches with the corresponding results in 
\cite{lisiatmupdate}. We have checked that at each C.L. the value of
the old and the new limit differs by this same factor.}. 
As a result of lowering of the
$\Delta m^2_{31}$ range according to the new 
SK atmospheric neutrino data analysis, 
the bound obtained from only CHOOZ data weakens:
as can be seen from figure \ref{delchith13},
this bound 
reads now as $\sin^2\theta_{13} < 0.096$. 
We have checked that the marginalised bounds on 
$\dm$ and $\sss$, even for the three-generation 
analysis, are the same as those given in Figure \ref{delvs12}.

In Figure \ref{solkl3osc} we present the allowed regions 
in the $\dm-\sss$ plane, for 
four fixed values of $\theta_{13}$ for cases with and without the 
SNO salt data. 
With the inclusion of the salt data the allowed regions decrease 
in size for all values of $\theta_{13}$. 
The presence of a small non-zero $\theta_{13}$ improves the fit
in the regions of 
the parameter space with higher values of $\dm$, 
i.e., in the high-LMA zone, where the $^8B$ neutrino transitions
are not affected by matter effects over a part of the energy 
spectrum \cite{Choubey:2001bi}, and which therefore 
give a larger value of the
CC/NC event rate ratio.
The presence of the $\cos^4\theta_{13}$ 
factor in the survival probability acts as 
a normalisation which effectively reduces
the CC event rate and hence makes the high-LMA region
less disfavored by the data. 
Since matter effects in the Sun are relatively small 
for the high-LMA values of $\dm$, 
the allowed regions in these zones appear at 
smaller values of $\theta_{12}$, as discussed above.
However, as $\theta_{13}$ increases, the allowed areas shrink and 
finally vanish for $\sin^2\theta_{13} > 0.075$.
Note that we get allowed regions at
$\sin^2\theta_{13}=0.075$ 
even though from Figure \ref{delchith13} 
the $3\sigma$ range appears to be $\sin^2\theta_{13} < 0.074$. This is 
because 
in Figure \ref{solkl3osc} we use a $\Delta \chi^2$ which corresponds 
to a three parameter fit.

\vspace{-0.4cm}
\section{Constraints on transitions into a 
state with a sterile neutrino component}
\vspace{-0.3cm}

As is well known, the 
explanation 
of the positive evidences of oscillations from 
the LSND  \cite{lsnd} experiment and
the solar and atmospheric neutrino oscillation data 
requires the existence of a fourth neutrino which has to be inert
\cite{giunti}.
A comparison of the SNO CC and NC 
data from the $D_2O$ phase 
had already ruled out solar $\nu_e$ oscillations
into pure sterile state at 5.3$\sigma$ \cite{Ahmad:2002jz}.
The inclusion of the SNO salt phase data
raises the degree of disfavour to 7.8$\sigma$ level 
\cite{Ahmed:2003kj}. 
However, transitions to ``mixed'' states, where 
the final neutrino state is a mixture of active and sterile 
components, is still allowed by the data. We find the limits 
on the sterile fraction from the global data. 
Bounds on allowed fraction of sterile component 
in the solar $\nu_e$ flux 
has earlier been obtained 
in \cite{sterilesolar}.   

\begin{figure}[t]
\centerline{\psfig{figure=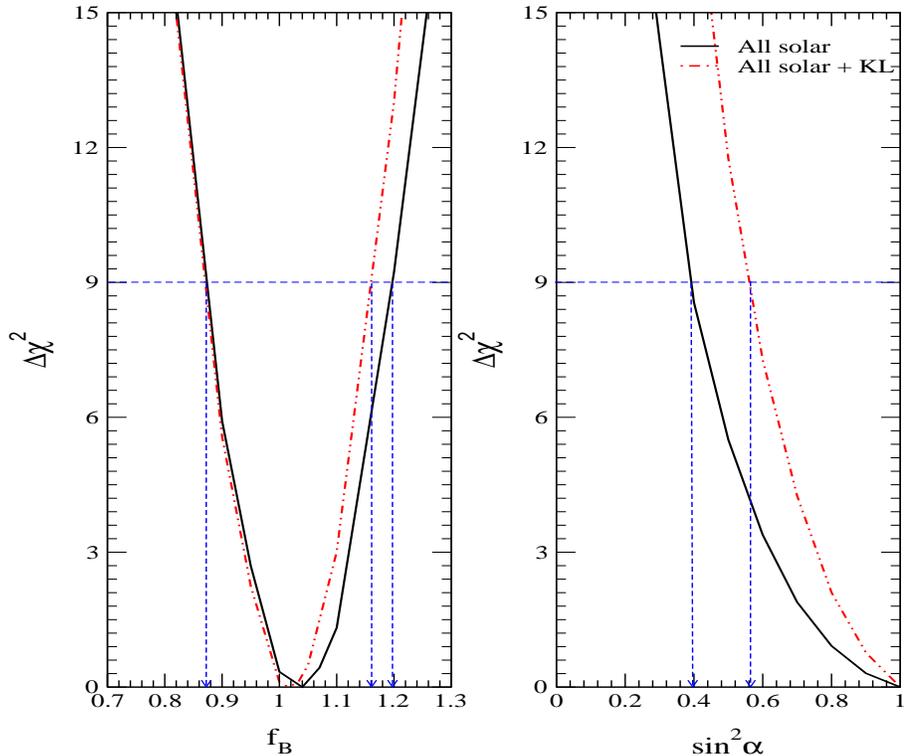,height=5.in,width=5.in}}
\caption{Bounds on $f_B$ and 
the sterile fraction in the solar neutrino flux, given by $1-\sin^2\alpha$. 
The left panel shows
the $\Delta \chi^2 $ as a function of $f_B$, while the right-hand panel 
gives the corresponding bounds on $\sin^2\alpha$. 
The $\Delta \chi^2 $ is marginalised over all the oscillation parameters.
The dashed line shows the $3\sigma$ limit corresponding to one parameter
fit.
}
\label{delchist}
\end{figure}

We consider a general case where the $\nue$ produced in the Sun 
transforms into a ``mixed'' state given by $\nu^\prime = 
\sin \alpha ~\nu_{active} + \cos \alpha ~\nu_{sterile}$. Thus, 
$\sin^2\alpha~(\cos^2\alpha)$ gives the fraction of the 
active~(sterile) component in the resultant solar neutrino 
flux at Earth.
In Figure \ref{delchist} we present the plots of 
$\Delta \chi^2$ vs $f_B$ (left-hand panel), 
and $\Delta \chi^2$ vs $\sin^2\alpha$ (right-hand panel), 
allowing the mass and mixing parameters to vary freely in the 
LMA region. The solid lines show the constraints 
from the solar data alone, 
while the dot-dashed lines correspond to the combined solar and \kl data.
The range of allowed values of $f_B$,
determined from the global solar~(solar+KamLAND) data analysis 
at $3\sigma$ is $0.87-1.2~(1.16)$. The allowed value 
for the sterile fraction in the resultant solar neutrino flux at
Earth is constrained to $1-\sin^2\alpha < 0.60~(0.44)$ at $3\sigma$
by the solar~(solar+KamLAND) data. 
Before the SNO salt phase data 
was announced, the corresponding limit 
for the sterile fraction at $3\sigma$ 
from the combined 
solar+KamLAND data analysis 
was $1-\sin^2\alpha < 0.54$.
Thus, the SNO salt data is seen to  
tighten the 
noose on the possible presence of a 
sterile component in the solar neutrino flux.

\vspace{-0.4cm}
\section{Future and outlook}
\vspace{-0.3cm}

\begin{figure}[t]
\centerline{\psfig{figure=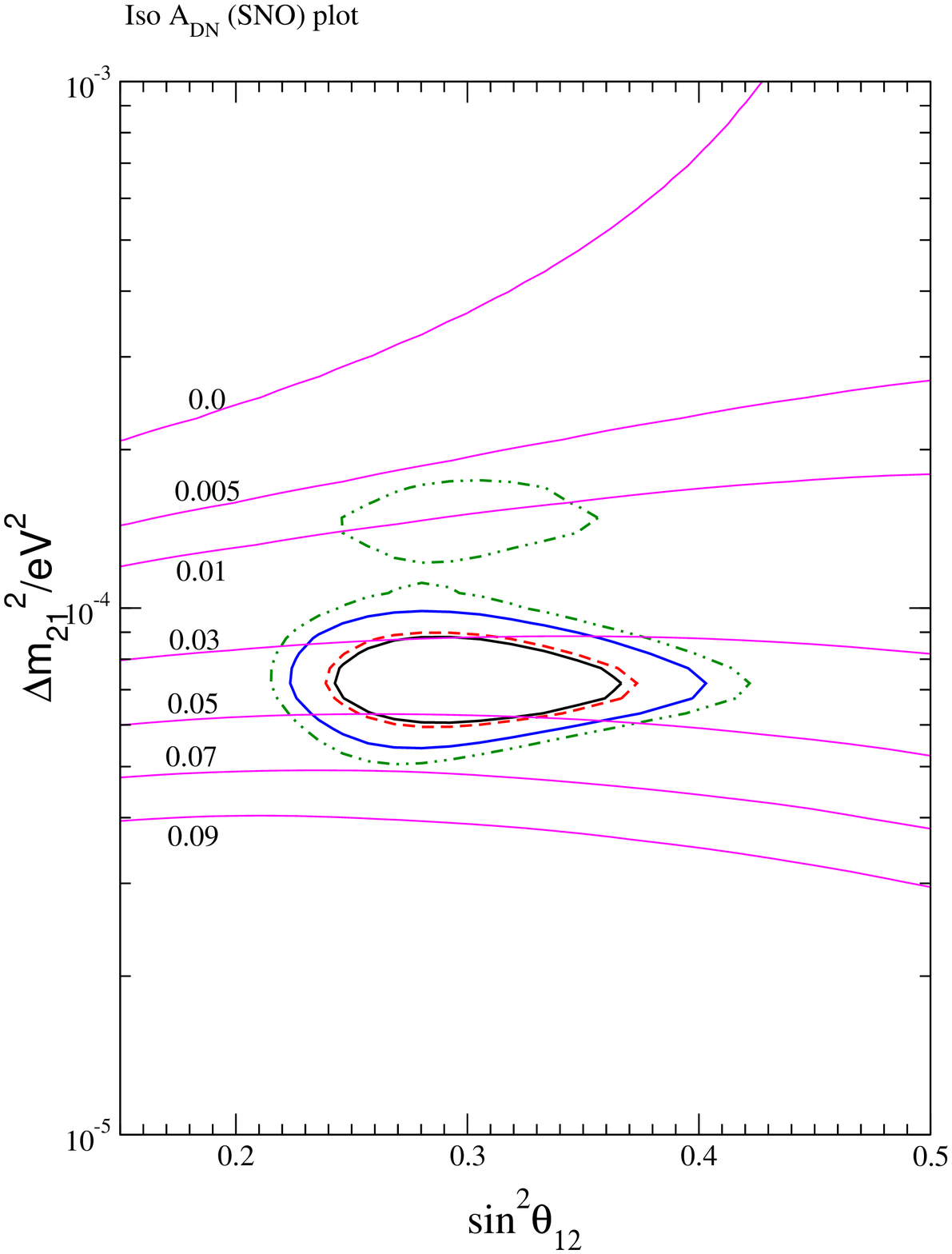,height=4.in,width=4.in}}
\caption{Lines of constant day-night asymmetry for the SNO
experiment, superposed on the allowed region from the global 
analysis of the solar and \kl data.} 
\label{dniso}
\end{figure}

\begin{figure}[t]
\centerline{\psfig{figure=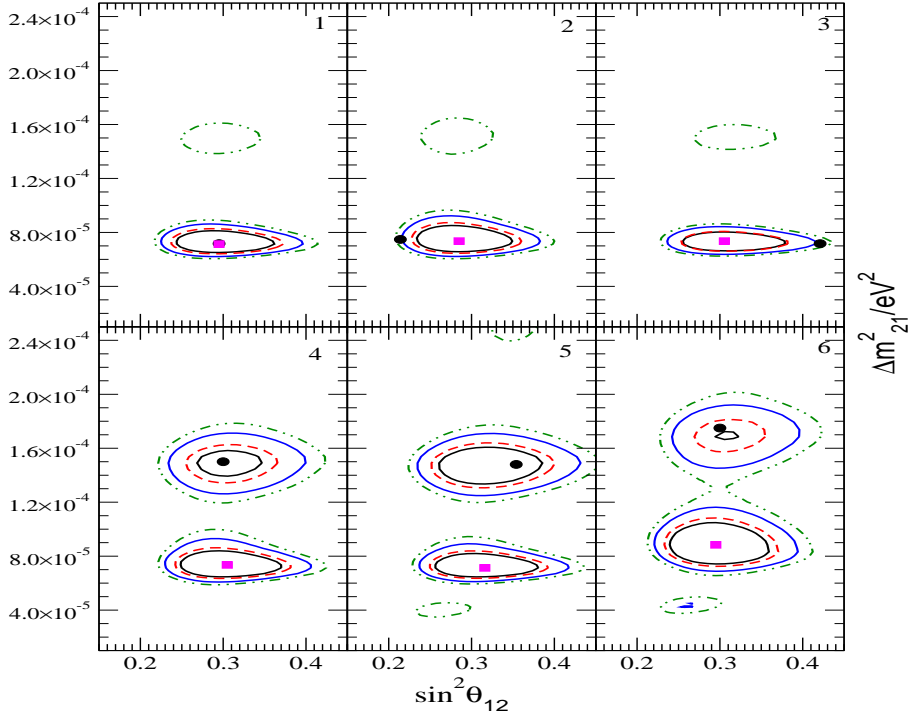,height=4.8in,width=5.in}}
\caption{
The 90\%, 95\%, 99\% and 99.73\% C.L. allowed regions obtained from a 
combined analysis using the global solar neutrino data and a 
0.41 kTy simulated KamLAND data. The points in the parameter 
space, for which the 0.41 kTy \kl data has been simulated, are shown 
by the dots; they have been chosen to lie 
within the current $3\sigma$ allowed regions.
The best-fit point of the combined analysis 
are shown as ``boxes''. 
}
\label{futuresolkl}
\end{figure}

\begin{figure}[t]
\centerline{\psfig{figure=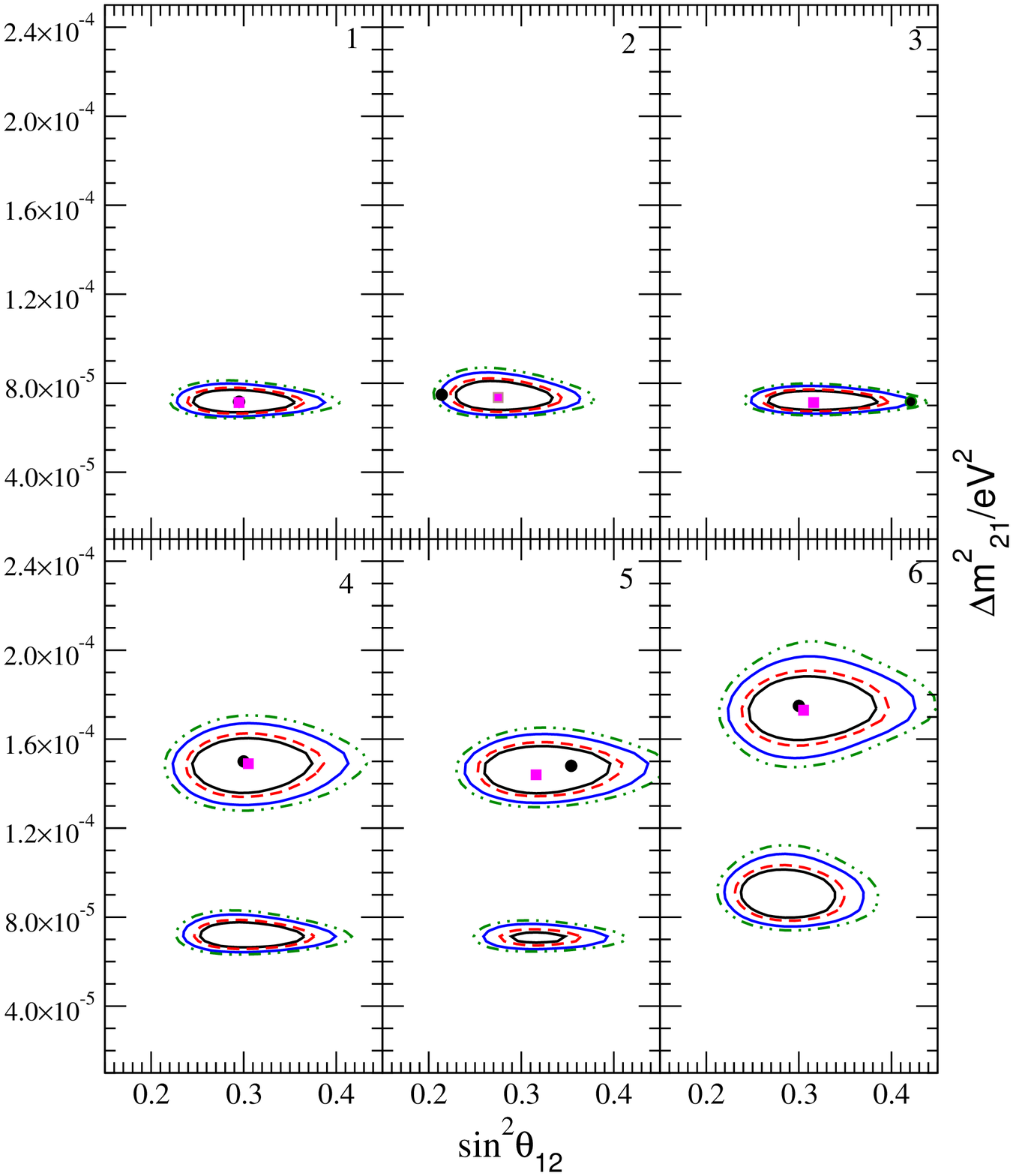,height=4.8in,width=5.in}}
\caption{Same as Fig. \ref{futuresolkl}, but for 1 kTy statistics.
}
\label{futuresolkl2}
\end{figure}

With the latest salt phase data from SNO 
giving further credence to the low-LMA solution, 
we are entering the era of precision
measurements in the field of 
solar neutrino physics. 
The high-LMA solution stands disfavored 
at more than 99\% C.L. 
and only a small area appears at the $3\sigma$ level.
The next phase of the
SNO experiment will be devoted to obtaining  
neutral current data using Helium counters \cite{sno3}. This 
would give a totally uncorrelated information on the CC and NC 
event rates observed at SNO. 
In the near future, SNO is expected to provide data on the 
day/night spectrum, which could be used in a statistical 
analysis to further constrain 
the solar neutrino oscillation 
parameters \cite{maris,maris2}. One of the related observables
is the day-night asymmetry:
\be
A_{DN} = 2\frac{N-D}{N+D}.
\ee
In Figure \ref{dniso} we show the lines 
of constant $A_{DN}$ for SNO.
The predicted $A_{DN}$ in SNO,
for the current best-fit values of the parameters  
in the low-LMA region, as well as the corresponding
$3\sigma$ range, are given by 
\be
A_{DN}^{SNO} = 0.04,~~3\sigma~ {\rm range}:~0.02-0.07,~~~{\rm low-LMA}, 
\ee
For the barely allowed high-LMA solution we get:  
\be
A_{DN}^{SNO} = 0.01,~~3\sigma~ {\rm range}:~0.007-0.02,~~~{\rm high-LMA}. 
\ee

The potential of Borexino \cite{borex} and any
generic 
electron scattering experiment for the low energy $pp$ neutrinos -- the 
LowNu experiments \cite{lownu} -- in constraining the 
mass and mixing parameters have been studied most recently in 
\cite{th12,Bahcall:2003ce}.
For the allowed regions obtained in the current paper, 
we find the predicted rates
for Borexino and LowNu experiments to be 
\be
R_{Be} &=& 0.65,~~(3\sigma~ {\rm range} \equiv 0.61-0.71);~~~{\rm low-LMA}\\
R_{pp} &=& 0.71,~~(3\sigma ~{\rm range} \equiv 0.67-0.76);~~~{\rm low-LMA}
\ee

High precision measurement 
results 
can be 
expected 
from the reactor experiment 
KamLAND \cite{preklall}. 
With statistics of 1 kTy,  
the KamLAND data 
could reduce the uncertainty in the 
\dm 
\\
determination to a 
few percent \cite{th12}. 
In Figures \ref{futuresolkl} 
and \ref{futuresolkl2}
we present 
the two-neutrino mixing 
allowed regions 
in the $\dm - \sss$ plane, 
obtained 
from a combined analysis of the 
current global solar neutrino data and a prospective 
0.41 kTy 
\footnote{This corresponds to 2.5 times the statistics of the 
first published data from KamLAND.} and
1 kTy simulated data in KamLAND. 
Since in future the systematic uncertainty in \kl data 
is expected to be reduced 
(especially with the fiducial volume calibration), 
we use a value of 5\% for the \kl systematic error in this analysis.
The black dots in the various 
panels of Figures \ref{futuresolkl} and \ref{futuresolkl2}
denote the point in the parameter 
space, for which the data has been simulated. The pink squares give the 
best-fit points obtained in the joint analysis. For the upper 
row of panels, the points at which we simulate the \kl data lie in the 
low-LMA region. 
We note that if the true solution lies in the low-LMA region, 
a spurious high-LMA solution still appears at $3\sigma$ level
 in the case of  0.41 kTy of statistics, 
though the allowed area gets further reduced in size, owing to the 
precision of the \kl data.
The high-LMA solution disappears if the statistics is 
increased to 1.0 kTy.
For all the panels the precision on the 
range of allowed value of \dm is seen to improve. 
There is little improvement, however, in the precision 
of $\sss$ \cite{th12,th12hlma}. 
One expects a significant improvement in the precision of $\sss$
over the low-LMA region to come from a more precise measurement \cite{taup}
of the CC/NC ratio at the next phase 
of SNO. 
In the lower row of Figures \ref{futuresolkl} 
and \ref{futuresolkl2}
we illustrate a 
scenario that would show itself if the future \kl data conforms to 
a point in the high-LMA region. Since the low-LMA solution 
is now strongly 
favored over the high-LMA one by the global solar neutrino data, if 
such a contradictory situation arises whereby the \kl data alone 
would favor the high-LMA solution,
both solutions would get allowed and the
solution ambiguity would remain.  
As can been seen in the last row panels of 
Figures \ref{futuresolkl} and \ref{futuresolkl2}, 
the best-fit point comes in the low-LMA (high-LMA) 
region for 0.41 kTy (1 kTy) statistics. 
We have checked that if we simulate the spectrum in the 
high-LMA region, the best-fit shifts from the low-LMA to 
the high-LMA region after \kl collects about 1 kTy statistics.

\vspace{-0.4cm}
\section{Conclusions} 
\vspace{-0.3cm}

We analysed the  impact of the
salt phase data from the SNO experiment 
in global solar neutrino 
oscillation analysis,
including the \kl data as well.
The inclusion of the CC and NC event 
rates from the SNO salt phase data
strongly favours
\dm to lie in the low-LMA region.
Values of \dm in the LOW area get
disfavoured at more than
3$\sigma$ just from global solar neutrino data, 
and at almost $5\sigma$ 
from the combined solar and \kl data.
The combined effect of the SNO spectrum 
data from the $D_2O$ phase
and of the data from the salt phase results in
lowering the upper bound on \dm to \dm $\leq 1.7\times10^{-4}$ eV$^2$
(99.73\% C.L.).
The global solar + \kl data still admit the high-LMA
solution, but it appears only at 2.63$\sigma$ level.
The addition of the new SNO 
data restricts the mixing angle $\theta_{12}$
from above and
maximal mixing is now excluded at more than 5$\sigma$.
With the inclusion of non-zero values of the mixing angle
$\theta_{13}$ in a 3-neutrino mixing analysis,
the allowed regions 
in the $\Delta m^2_{21} - \sin^2\theta_{12}$ plane
decrease in size as $\theta_{13}$ increases.
At $\sin^2\theta_{13} \geq 0.075$ no allowed 
regions are obtained at 99.73\% C.L.
The solution due to transitions into sterile neutrino
is excluded  at 7.8$\sigma$ with the salt phase data.
However, solar $\nu_e$ transitions into a mixed sterile + active 
state are allowed,  
with the sterile fraction 
restricted to be $<$ 44\% at 3$\sigma$.
With the knowledge of   
$\Delta m^2_{21}$ and  $\sin^2\theta_{12}$
responsible for the solar neutrino oscillations
becoming more precise,
the predicted ranges for the
day-night asymmetry in SNO, and of the 
event rates in Borexino
and the LowNu, experiments narrow down.
We also studied the impact of the prospective increase in statistics of 
the \kl data, 
on the determination of the solar neutrino oscillation 
parameters. If the spectrum is simulated at a point
in the low-LMA region, 
the allowed 3$\sigma$ area in the 
high-LMA zone reduces in size 
in the case of 0.41 kTy of data, and disappears if
the statistics is increased to 1.0 kTy.
If, however, the 
\kl spectrum corresponds to a point in the high-LMA zone,
the conflicting
trend of solar and \kl data would make the high-LMA
solution reappear at 90\% C.L. and the determination of \dm would
remain ambiguous.

\vskip 8pt
The authors would like to thank E. Lisi  
for clarifying comments concerning ref. \cite{lisiatmupdate}.
S.G. would like to thank The Abdus Salam International Centre for
Theoretical Physics for hospitality.

\vspace{-0.3cm}


\end{document}